# Budget Allocation for Power Networks Reliability Improvement: Game-Theoretic Approach

Hamzeh Davarikia
Electrical and Computer Engineering Department
Louisiana State University
Baton Rouge, LA, USA

Masoud Barati
Electrical and Computer Engineering Department
University of Pittsburgh
Pittsburgh, PA, USA

Yupo Chan
Systems Engineering Department
University of Arkansas at Little Rock
Little Rock, AR, USA

Kamran Iqbal
Systems Engineering Department
University of Arkansas at Little Rock
Little Rock, AR, USA

*Abstract*—Budget allocation for power system reliability improvement is considered among the sophisticated problems because of its nonlinear nature. This nonlinearity makes the problem intractable for large-scale power systems. This paper compares two approaches for budget allocation for power system reliability improvement. The first method is the traditional one, which suffers from the non-convexity and nonlinearity. In the second approach, a linear zero-sum mixed-strategy game is proposed where a limited budget is allocated among the network elements based on the game variables along with an iterative algorithm to find the solution. Both models are applied to the modified RTBS system. The results show that while each strategy adopts a different tactic for reliability improvement, both strategies improve the system reliability to the same level for a given budget.

## I. INTRODUCTION

Electric power network is among the most sophisticated systems that have been evolved by engineering design. The power system is always encountered the risks, including the physical intrusion and the natural disasters [1], treats imposed from the nature of the power networks like the dynamic instability [2]-[4], oscillation of the grid-connected renewable generation resources [5], [6] or other sort of components that are always threaten the networks [7]-[11]. Different approaches including data security enhancement [12], [13], optimization, operational procedure and design of new approaches to mitigate and protect counter measures are used to improve the reliability and resiliency of the power grids [14]-[22]. Innovative and effective protective algorithm and damping controller is proposed to mitigate the oscillation of renewable generation resources in the vicinity of series compensated lines in [6]. While there is a wealth of research dealing with optimization in grid operation through the stochastic [14] or deterministic approaches [15]-[18], but there is a little attraction to improve power system resilience, or the reliability through the optimization approaches. Salehi et. al [19] evaluated the automated substation reliability with different configuration. The power system resilience improvement sought through network reconfiguration in [20], which can be considered as a cost-effective redeem for bulk power system. This approach can be utilized in case of extreme event or crafted attack to facilities the power restoration [22] as a cost-effective approach.

Improving power systems reliability against such risks can be achieved through the resource allocation models, which seek to maximize statistical-reliability of the system [23]. The system's statistical-reliability is the probability that a source-to-sink path exists in a system at a particular time instant, given the network is only subject to statistical failures [24]. The interpretation of this concept in the power systems context is to define the probabilities that alternate paths between loads and generators are available. To this end, we developed two optimization models where network planner can recognize the most invulnerable elements and allocate his limited resource in the right place to improve network reliability.

Similar to our work, authors in [25] consider the power network as the extensive system consisting of subsystems connected in series and parallel. Power system reliability improvement sought through forming trees of power flow paths, which facilitate improving subsystem's reliability. An integrated optimization and simulation method is proposed in [26] that nominates the facilities for the reliability improvement considering the limited budget. The elaborate Monte Carlo simulation approach along with an instance of the Genetic Algorithm makes their model computationally expensive, particularly in the case of large-scale power networks.

A nonlinear approach to power system reliability indices improvement is discussed in [27]. While the nonlinearity is the most prominent challenge of their model, ant colony optimization method is employed to find solutions. A two-player non-cooperative game is proposed by [28], where the router and attacker are the players. While the attacker targets the most critical link in the network, a router seeks to minimize the system cost by changing the operation strategies. Similarly, [29] proposed a game-theoretic approach, where the author recognizes the essential power network's facilities, which are vulnerable to attack.

While investment planning against deliberate attack is quite essential and attracts a considerable amount of attention among researchers [29], there is a dearth of work in investment planning against natural hazards in the power network, which happen frequently [30]. In this paper, we propose models for improving network resiliency to address the natural hazards problems. In particular, our paper considers two analytical models: one, a traditional improvement model, and two, a game-theoretic model. In the first approach, a nonlinear model is proposed to allocate the limited budget to the elements to maximize power system statistical reliability. While this method can distribute the budget efficiently among the components, the nonlinearity results in an inefficient (or perhaps intractable) model for the large-scale power network.

The second approach is a linear programming mixed-strategy zero-sum game, which assigns a portion of the budget to the most vulnerable elements with an iterative algorithm. The algorithm will continue until the budget is exhausted or system reaches the desired reliability level. The modified RTBS test system is employed to test the model's performance, and the results illustrate that with the same budget, the two approaches improve the network reliability to the same level, albeit with different strategies. The rest of this paper organized as follows: section II proposes the traditional reliability improvement model and explains the model with an illustrative example. Section III discusses the game-theoretic model, forming the payoffs matrix, the steps of algorithms and presents the case study. Finally, Section IV gives the conclusion.

## II. TRADITIONAL RELIABILITY IMPROVEMENT MODEL

The traditional nonlinear reliability improvement model is proposed in this section. Fig. 1 shows the modified RTBS system [31], where the reliability of generators and lines demonstrated beside each element. This grid can be interpreted as the network with multiple origins (O) and multiple destinations (D), or Multi OD network in short. In this case, the sources are generators G1 and G2, and the destinations are loads L1-L4, and power traverses the paths between each OD.

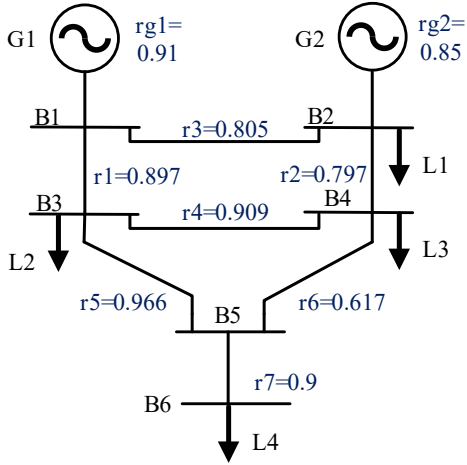

Fig. 1. Modified RTBS system

To calculate the statistical reliability between generators and lodes, the possible routes between ODs is first enumerated. Since the paths between an OD are in parallel, the OD reliability of the system with n paths are calculated with the following equation:

$$R_{O-D} = 1 - (1-p_1)(1-p_2)...(1-p_n) \quad (1)$$

where $p_n$ is the reliability of path $n$. For example, the paths between G1 and L4 are:

$Path\ 1 = [1,2,4,3,5,6], p_1 = r_{g1}r_3r_2r_4r_5r_7 = 0.461$
$Path\ 2 = [1,2,4,5,6], p_2 = r_{g1}r_3r_2r_6r_7 = 0.324$
$Path\ 3 = [1,3,4,5,6], p_3 = r_{g1}r_1r_4r_6r_7 = 0.412$
$Path\ 4 = [1,3,5,6], p_4 = r_{g1}r_1r_5r_7 = 0.71$

Then, according to (1), reliability between generator 1 and load 4 is:

$$R_{G_1-L_4} = 1 - (1-r_{g1}r_3r_2r_4r_5r_7)(1-r_{g1}r_3r_2r_6r_7)$$
$$(1-r_{g1}r_1r_4r_6r_7)(1-r_{g1}r_1r_5r_7) = 0.938$$

As can be seen in the abovementioned simple calculations, power can traverse through four different paths from G1 to L4. While each track is not highly reliable, the $R_{G_1-L_4}$ is relatively reliable because the paths are in parallel. Based on this concept, we propose a resource allocation based Multi OD reliability improvement model (2)-(6) where system planner can allocate his limited budget to improve power network resiliency against natural hazards and hence improve system reliability.

$$\max \frac{1}{N_{od}} \sum_{o \in G} \sum_{d \in L} R_{od} \quad (2)$$

subject to

$$r_i = r_i^0 + x_i; \quad \forall i \in E \quad (3)$$

$$\sum_{i \in E} c_i x_i = B^R \quad (4)$$

$$0 \leq r_i \leq 1; \quad \forall i \in E \quad (5)$$

$$R_{od} = 1 - \prod_{p \in Paths_{od}} (1-R_{od}^p); \forall o \in G, \forall d \in L \quad (6)$$

In the above equations $N_{od}$ is the number of total ODs (8 in our case study), $R_{od}$ is the statistical-reliability between origin $o$ and destination $d$, $r_i$ is the components reliability after improvement $x_i$ added to the initial reliability $r_i^0$, $c_i$ is the cost of reliability improvement for element $i$, which are considered as 1 for the lines and 2 for the generators, $Paths_{od}$ is the set of paths between origin $o$ and destination $d$, $R_{od}^p$ is the reliability of path $p$ between origin $o$ and destination $d$, $E$ is the set of system's components, $G$ is the set of generators or origins, $L$ is the set of loads or destinations, and $B$ is the total budget for reliability improvement..

In the traditional model, the nonlinear equation (6) is used to calculate reliability between each load-generation (OD), while (2) serves as an index to measure the overall reliability of the power system. Further, (3) is the reliability improvement constraint, and (4) is the budget constraint for the reliability improvement. Finally, the reliability of each (OD) is ensured to be in 0-1 range by (5).

The modified RTBS reliability test system shown in Fig. 1 is employed in this section to show the model's performance. The first step is to enumerate all the paths between generators and loads. Table I shows the paths between generators and loads in the RTBS system, which forms the components of sets

$Paths_{od}$. Table II displays the statistical-reliability between generators and loads in each path shown in Table I, and Table III demonstrates the overall reliability of each load and generator before improvement, or when $B^R = 0$. The objective function before improvement is 0.917. This is the normalized summation of all ODs reliability, which is an 0-1 ranged index and stands for the power system reliability.

TABLE I. PATHS BETWEEN GENERATORS AND LOADS

|  |  | Loads (destinations) | | | |
|---|---|---|---|---|---|
|  |  | L1 | L2 | L3 | L4 |
| Generators | G1 | {1,2}<br>{1,3,4,2}<br>{1,3,5,4,2} | {1,2,4,3}<br>{1,2,4,5,3}<br>{1,3} | {1,2,4}<br>{1,3,4}<br>{1,3,5,4} | {1,2,4,3,5,6}<br>{1,2,4,5,6}<br>{1,3,4,5,6}<br>{1,3,5,6} |
|  | G2 | {2} | {2,1,3}<br>{2,4,3}<br>{2,4,5,3} | {2,1,3,4}<br>{2,1,3,5,4}<br>{2,4} | {2,1,3,4,5,6}<br>{2,1,3,5,6}<br>{2,4,3,5,6}<br>{2,4,5,6} |

TABLE II. STATISTICAL-RELIABILITY OF OD PATHS - BEFORE IMPROVEMENT

| Generator | Load | $p_1$ | $p_2$ | $p_3$ | $p_4$ |
|---|---|---|---|---|---|
| G1 | L1 | 0.733 | 0.591 | 0.388 | |
|  | L2 | 0.816 | 0.531 | 0.348 | |
|  | L3 | 0.584 | 0.742 | 0.487 | |
|  | L4 | 0.461 | 0.324 | 0.412 | 0.71 |
| G2 | L1 | 0.85 | | | |
|  | L2 | 0.614 | 0.616 | 0.404 | |
|  | L3 | 0.677 | 0.558 | 0.366 | |
|  | L4 | 0.31 | 0.534 | 0.535 | 0.376 |

TABLE III. OD RELIABILITY - BEFORE IMPROVEMENT

|  | L1 | L2 | L3 | L4 |
|---|---|---|---|---|
| G1 | 0.933 | 0.944 | 0.945 | 0.938 |
| G2 | 0.85 | 0.912 | 0.91 | 0.907 |

To examine the model performance, reliability budget $B^R$ is gradually increased from 0 to 1 with an interval of 0.1. Fig. 2 depicts the trend of system reliability improvement when the budget is increasing. As can be seen, the system reliability increased considerably when the budget is ranging from 0 to 0.4. Fig. 3 shows sensitivity of each OD pair to the budget increment. In this figure, the sharpest reliability increment belongs to the $R_{G_2-L_1}$ where there is only one element on its path. While each path has its own trend, the interesting characteristic of this figure is that the reliability of the paths that originate from one generator has almost the same trend.

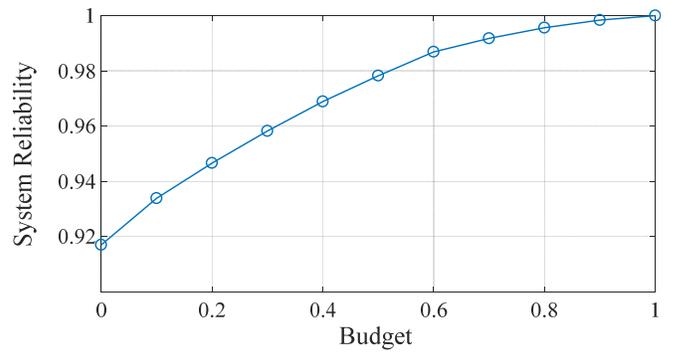

Fig. 2. Sensitivity of system's reliability to budget increment

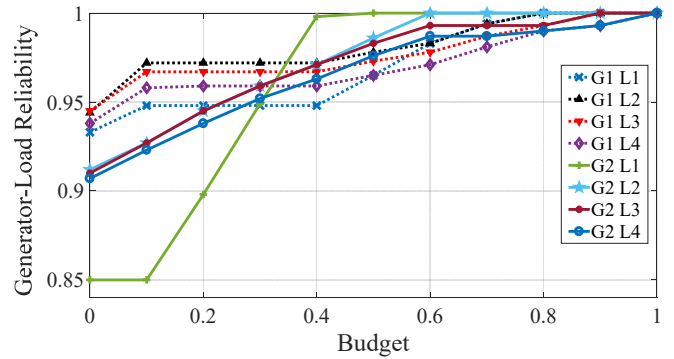

Fig. 3. Sensitivity of OD reliabilities to budget increment

Fig. 4 shows the reliability improvement for each element in a different budget. As can be seen $r_1$, $r_{g1}$ $r_{g2}$ and $r_3$ got the reliability improvement in the various amount of budget. One can say these components are among the important elements within the system. Interestingly, $r_6$, which has the lowest reliability in the system did not receive any improvement through the proposed model.

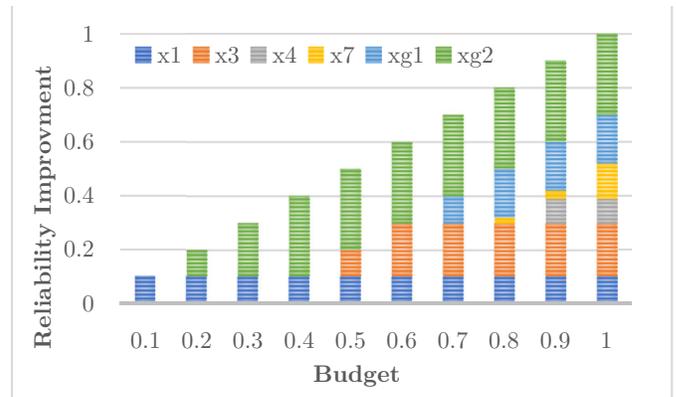

Fig. 4. Element's reliability improvement in different budget

After this illustrative example, the next section proposes a linear game-theoretic model for reliability budget allocation model

## III. RELIABILITY IMPROVEMENT MODEL: GAME-THEORETIC APPROACH

While the model proposed in Section II allocates the budget to the proper component, the nonlinearity of (6) may cause an intractable problem for the large-scale power network. In this section, a game-theoretic approach is proposed which helps the system planner to recognize the most vulnerable elements against natural hazards. To this end, a two-player zero-sum mixed-strategy game between "Blue" (network planner) and "Red" (Mother Nature) is established here, where Blue can improve network resiliency with spending budget and players can only target one component at a time.

Suppose that the overall network statistical reliability before improvement is $R^0$. If Red destroys element, the statistical reliability will decrease to $R'_i$, and the associated damage to the system reliability is $(R^0 - R'_i)$. This damage is ranging from 0 to 1 since the $R'_i$ is less than or equal to $R^0$. In order to improve system resiliency, we need a function that may evaluate the invulnerability of the system. This can be achieved by defining damage utility function as $y_i = 1 - (R^0 - R'_i)$, which is a 0-1 ranged function that measures system invulnerability resulting from a Blue decision.

The game payoffs, depicted in Fig. 5, shows the outcome of a pure strategy game, i.e., the payoff allowed when Red and Blue each pick a specific row and column strategy respectively. Let us introduce the game variable $\psi_i$ as the percentage of total portion of budget consumed on component $i$ in a game. The percentage of time Blue should adopt the strategy $\psi_i$ to improve reliability of component $i$ is attributed to a row of the game matrix. In the payoff matrix, each row is the Blue's strategy, while each column indicates Red's decisions. Blue protects and Red tampers the component $i$ in the $i^{th}$ row and $i^{th}$ column respectively. Accordingly, the diagonal elements are those for which Blue has already made hundred percent reliability improvement strategies and hence they are totally invulnerable to damage, while the off-diagonal elements in column $i$ indicate the damage utility function when Red tampers component $i$.

|  | $r_1$ | $r_2$ | $r_3$ | $r_4$ | $r_5$ | $r_6$ | $r_7$ | $r_{g1}$ | $r_{g2}$ |
|---|---|---|---|---|---|---|---|---|---|
| $r_1$ | 1 | $y_2$ | $y_3$ | $y_4$ | $y_5$ | $y_6$ | $y_7$ | $y_{g1}$ | $y_{g2}$ |
| $r_2$ | $y_1$ | 1 | $y_3$ | $y_4$ | $y_5$ | $y_6$ | $y_7$ | $y_{g1}$ | $y_{g2}$ |
| $r_3$ | $y_1$ | $y_2$ | 1 | $y_4$ | $y_5$ | $y_6$ | $y_7$ | $y_{g1}$ | $y_{g2}$ |
| $r_4$ | $y_1$ | $y_2$ | $y_3$ | 1 | $y_5$ | $y_6$ | $y_7$ | $y_{g1}$ | $y_{g2}$ |
| $r_5$ | $y_1$ | $y_2$ | $y_3$ | $y_4$ | 1 | $y_6$ | $y_7$ | $y_{g1}$ | $y_{g2}$ |
| $r_6$ | $y_1$ | $y_2$ | $y_3$ | $y_4$ | $y_5$ | 1 | $y_7$ | $y_{g1}$ | $y_{g2}$ |
| $r_7$ | $y_1$ | $y_2$ | $y_3$ | $y_4$ | $y_5$ | $y_6$ | 1 | $y_{g1}$ | $y_{g2}$ |
| $r_{g1}$ | $y_1$ | $y_2$ | $y_3$ | $y_4$ | $y_5$ | $y_6$ | $y_7$ | 1 | $y_{g2}$ |
| $r_{g2}$ | $y_1$ | $y_2$ | $y_3$ | $y_4$ | $y_5$ | $y_6$ | $y_7$ | $y_{g1}$ | 1 |

Fig. 5. Payoff matrix for game-theoretic model

From the game matrix, the game value can be obtained by calculating the expected value of each Red strategy. For example, the expected value for the Red's tampers on element $r_2$, corresponding to column 2, is $y_2(\psi_1 + \psi_3 + \psi_4 + \psi_5 + \psi_6 + \psi_7 + \psi_{g1} + \psi_{g2}) + \psi_2$. Based on the above explanation, (7)-(9) shows the game-theoretic model, where $v$ is the value of the game, which is equal to statistical-reliability of the system.

$$\max v \quad (7)$$

subject to

$$v \leq y_i \sum_{\substack{i,j \in E \\ i \neq j}} \psi_j + \psi_i; \quad \forall i \in E \quad (8)$$

$$\sum_{i \in E} \psi_i = 1 \quad (9)$$

Constraint (8) is the game constraints as explained above and (9) enforces the game variables sum to 1, promising a mixed-strategy defensive game. The proposed game should repeat iteratively to allocate the total budget to the elements or in order to reach the desired reliability level. To this end, our proposed algorithm is described as per the following steps.

***Step 0***: set the iteration counter $t = 0$.

***Step 1***: $r_i \leftarrow r_i^t$. Calculate system reliability using (2) and obtain payoff matrix $y^t$. Run the LP model (7)-(9) to obtain $\psi_i^t$.

***Step 2***: $t = t + 1$. Pump the budget $B^t$ until the first element reaches to perfect reliability 1.00 using the following equation: $r_i^t = \psi_i^{t-1} B^t / c_i + r_i^{t-1}$.

***Step 3***: calculate the remaining budget by $B^R - \sum B^t$ If the total budget $B^R$ is not exhausted or the desire system reliability is not achieved, go to ***Step 1***, otherwise return the system reliability $r_i \leftarrow r_i^t$.

The modified RTBS system is employed to demonstrate the algorithm performance. The target is to have perfect reliability 1.00 based on (2). Table IV shows one iteration of the proposed model. The column $r_i^0$ simply displays the initial elements reliability as shown in Fig.1. Column $R'^0_i$ shows the reliability degradation as a result of damaging component $i$, which is needed to calculate damage utility function shown in column $y_i^0$. According to ***Step 1***, $\psi_i^0$ are the game variables obtained from LP problem (7)-(9). Now we can pump the budget $B^1$ based on the equation shown in ***Step 2***. To this end, the budget is increased until the first element reaches its desired reliability. In this case, by increasing the budget to $B^1 = 0.35$, the $r_{g1}$ is the first elements that reaches to the perfect reliability of 1.00, meanwhile there are some reliability improvement for $r_7$ and $r_{g2}$ based on their game variable. At the end of iteration 1, the system reliability jumped from 0.917 to 0.952. This procedure is repeated six times, consuming total budget 1.00 and the system reliability (2) reaches to the perfect value of 1.00.

Fig. 6 shows the cumulative reliability improvement for the six iterations. As an instance, while the second iteration allocates the budget equal to 0.205, the associated budget in the figure is 0.555, which is summation of 0.205 in the second iteration and 0.35 in the first iteration. As can be seen, while the budget and system reliability target are similar to the case shown in Fig.4, the game-theoretic model employed a different strategy rather than the traditional model. The improvement results are clearly visible in Fig.7, where a comparison between both models is provided for each element.

TABLE IV. INSTANCE OF ONE ITERATION OF THE PROPOSED METHOD

| Element | $r_i^0$ | $R_i'^0$ | $y_i^0$ | $\psi_i^0$ | $r_i^1$ |
|---|---|---|---|---|---|
| $r_1$ | 0.897 | 0.707 | 0.79 | 0 | 0.897 |
| $r_2$ | 0.797 | 0.763 | 0.846 | 0 | 0.797 |
| $r_3$ | 0.805 | 0.784 | 0.867 | 0 | 0.805 |
| $r_4$ | 0.909 | 0.804 | 0.887 | 0 | 0.909 |
| $r_5$ | 0.966 | 0.804 | 0.887 | 0 | 0.966 |
| $r_6$ | 0.617 | 0.86 | 0.943 | 0 | 0.617 |
| $r_7$ | 0.9 | 0.687 | 0.77 | 0.002 | 0.901 |
| $r_{g1}$ | 0.91 | 0.447 | 0.53 | 0.512 | 1.000 |
| $r_{g2}$ | 0.85 | 0.47 | 0.553 | 0.486 | 0.935 |

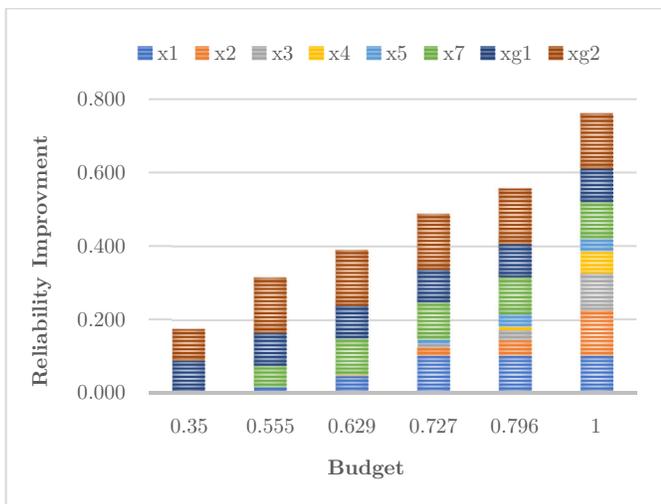

Fig. 6. Inctance of one iteration of the proposed method

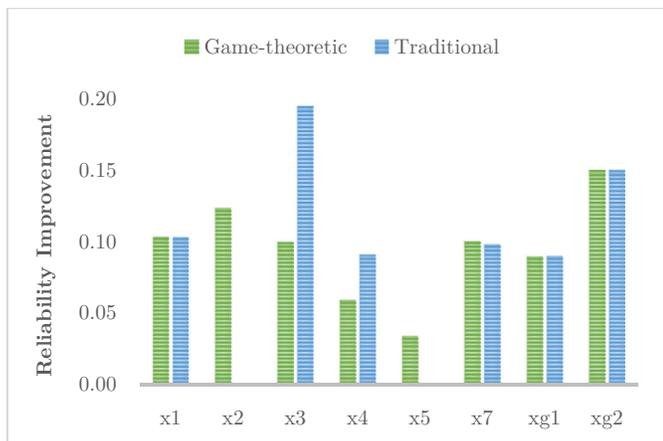

Fig. 7. Inctance of one iteration of the proposed method

IV. CONCLUSION

Power system reliability improvement is among the most sophisticated problems because of its nonlinear nature. This nonlinearity makes the problem considerably complicated for large-scale power systems. We define the reliability as the availability of paths between loads and generators and propose an index for overall system's reliability based on the normalized summation of all the routes between demands and sources. Then we introduce two budget allocation models for power grid reliability improvement, where the first model follows traditional approach to reliability improvement and the second model is a game-theoretic linear programming model. In the first model, the limited budget is allocated to the elements to maximize system reliability by nonlinear programming. However, this approach is highly non-convex and non-linear for large-scale problems, which commonly result in an inefficient solution.

In the second method, we propose a linear two-player zero-sum mixed-strategy game for the budget allocation problem. This game should repeat iteratively using the proposed algorithm until the budget is exhausted, or the reliability reaches the desired level. The second approach is linear, and hence convex, and promotes effective solutions even for the large-scale power systems. The case studies run on the modified RTBS systems shows the two approaches result in the same reliability level for a given budget, albeit with a different strategy. For the future work, one may apply the model to a large-scale power network as well as further improve the algorithm used in the game-theoretic approach.

Moreover, Mother Nature causes damage to an electric grid as a matter of randomness. She is not an adversary and does not cause damage deliberately. The zero-sum game used in this paper treated her as an adversary, which is a conservative consideration. One may address this issue using other than a zero-sum game like a non-cooperative game, which will fit Mother Nature role more accurately.